\documentclass[preprint,showpacs,preprintnumbers,amsmath,amssymb]{revtex4}

\usepackage{graphicx}
\usepackage{dcolumn}
\usepackage{bm}
\newcommand{\beq} {\begin{equation}  }
\newcommand{\eeq} {\end{equation}    }
\newcommand{\bea} {\begin{eqnarray}  }
\newcommand{\eea} {\end{eqnarray}    }
\newcommand{\M}{\,{\mathcal M}}
\newcommand{\lm}{\lambda}
\newcommand{\TeV}{\,{\rm TeV}}
\newcommand{\GeV}{\,{\rm GeV}}
\newcommand{\lsim}{\mathrel{\mathop{\kern 0pt \rlap
  {\raise.2ex\hbox{$<$}}}
  \lower.9ex\hbox{\kern-.190em $\sim$}}}
\newcommand{\gsim}{\mathrel{\mathop{\kern 0pt \rlap
  {\raise.2ex\hbox{$>$}}}
  \lower.9ex\hbox{\kern-.190em $\sim$}}}

\begin{document}

\preprint{hep-ph 0306112 ~~~\\ KIAS P03039}

\title{
Muon anomalous magnetic moment\\
and the heavy photon in a little Higgs model}

\author{Seong Chan Park}
 \email{spark@kias.re.kr}
\affiliation{%
Korea Institute for Advanced Study (KIAS)\\
    207-43 Cheongryangri-dong Dongdaemun-gu, Seoul 130-012, Korea
}%
\author{Jeonghyeon Song}%
 \email{jhsong@konkuk.ac.kr}
\affiliation{%
Department of Physics, Konkuk University,
                   Seoul 143-701, Korea
}
\date{\today}

\begin{abstract}
In the Littlest Higgs model, we comprehensively study the
phenomenology of the heavy photon $A_H$ which is lightest, in most
of the parameter space, among newly introduced heavy gauge bosons
and top-like vector quark. Unexpected behavior is that lighter
$A_H$ suppresses the corrections to the electroweak precision
observables. For the global symmetry breaking scale $f \lsim 3$
TeV, the heavy photon can be light enough to be produced at the
$500 \GeV$ linear collider. Through the calculation of the
one-loop correction to the muon anomalous magnetic moment, we show
that even the light $A_H$ with mass around 200 GeV results in the
negligible contribution. This is consistent with the current
inconclusive status of the theoretical calculation of the
$(g-2)_\mu$ in the SM. The effects of the Littlest Higgs model on
the process $e^+ e^- \to \mu^+ \mu^-$ are also studied, which is
one of the most efficient signals to probe the $A_H$.
\end{abstract}

\maketitle

\section{Introduction}
\label{SecIntroduction}

The standard model (SM) of particle physics has provided an
excellent effective field theory of high energy phenomena up to
the energies of order 100 GeV.
A direct and important question is what is the cutoff scale of
this effective description? The Higgs mass of the SM may have a
key because of its quadratic sensitivity to heavy physics. The
naturalness argument suggests that the cutoff scale of the SM
cannot be too higher than the electroweak scale: New physics will
appear around $\TeV$ energies. A weak scale supersymmetry (SUSY)
model is one of the best motivated candidates for new physics as
the cutoff scale is naturally replaced by the soft SUSY
breaking scale.
Even though various supersymmetry models have been thoroughly
studied by enormous number of authors, no piece of experimental
data has been discovered. Brane world scenarios with large or
warped extra dimensions have been suggested to understand the
hierarchy problem as a geometrical stabilization problem. However
those theories are not weakly coupled at TeV scale.

Recently, new models, dubbed the ``little Higgs'' models, have
drawn a lot of interest, which remain weakly coupled at TeV
sale with the one-loop stabilized Higgs potential. The original
idea dates back to
1970's, such that the lightness of the Higgs boson is attributed to its
being a pseudo Goldstone boson\,\cite{Georgi:yw}.
The problem was then the remaining
quadratic divergence of the radiative correction to the Higgs
mass. A new ingredient, the collective symmetry breaking, was
discovered via dimensional deconstruction\,\cite{d1,d2}. It
ensures that the Higgs mass is radiatively generated at two loops
\footnote{See Ref.~\cite{rv1}\cite{rv2} and \cite{rv3} for
pedagogical reviews.}. Phenomenologically the quadratic
divergences due to the SM gauge bosons (top quark) are cancelled
by those due to new heavy gauge bosons (fermions): The
cancellation occurs at one-loop level between particles with the
same statistics, unlike the cancellations in supersymmetric
theories; it is due to the exactly opposite coupling of the
new particles compared to the SM particles. The heavy gauge bosons
come from the extended gauge sector which fixes the new gauge
coupling structures. In addition, the requirement of cancellation
of quadratic divergence fully determines the Yukawa sector of the
top quark. Later the idea has been realized in other simple
nonlinear sigma models\,\cite{minimal moose,littlest
higgs,sp6,simple group,custodial su2,SO9,SU9}\footnote{It is a
challenging problem to obtain UV-complete theory. See papers by
Kane\,\cite{kane} and Nelson\,\cite{nelson} for discussions.}.

In this paper, we concentrate on ``the Littlest Higgs'' model,
described by the global symmetry breaking pattern
$SU(5)/SO(5)$\,\cite{littlest higgs}. As one of the simplest
realizations of the little Higgs idea, it is
the smallest extension of the SM to date 
which stabilizes the electroweak scale and remains weakly coupled
at TeV scale. The model predicts the presence of new heavy gauge
bosons ($W_H$, $Z_H$ and $A_H$) and a new heavy top-like vector
quark $T$ and their couplings. The minimality of the Littlest
Higgs model would leave characteristic signatures at the present
and future collider experiments. Since the tree level corrections
of the Littlest Higgs model to electroweak precision data
constrain the heavy particles as massive as a few TeV, a 500 GeV
linear collider (LC) has not been expected to efficiently test the
model. In literatures, Large Hadronic Collider(LHC) of the CERN is
shown to have a potential to detect the new
particles\,\cite{c1,c2,c3}. In the Littlest Higgs model, however,
we find that the global symmetry structure $SU(5)/SO(5)$ yields
substantially light $A_H$, light enough to be produced on-shell at
a 500 GeV LC. Moreover, as shall been shown below, the $A_H$
becomes lighter in the parameter space where the corrections to
the electroweak precision measurements are minimized.

The presence of a few hundreds GeV heavy photon can be dangerous
to other low energy observables. We study its one-loop
contributions to a well measured observable, the muon anomalous
magnetic moment. Another issue here is the collider signatures of
the $A_H$. The process of $e^+ e^- \to \mu^+ \mu^-$ is to be
discussed, which is one of the most effective processes to probe
the model, as the branching ratios (BR) of the heavy photon
suggests.

The paper is organized as follows. In Sec.~\ref{SecLH}, we briefly
review the Littlest Higgs model. We point out the preferred
parameter space by considering some tree level relations of the SM
gauge boson masses and couplings. In Sec.~\ref{SecAH}, physical
properties of the heavy photon is studied, focused on its mass and
decay patterns. In Sec.~\ref{SecMuon}, the one-loop corrections of
the new gauge bosons to the muon anomalous magnetic moment are
calculated. Numerical value is to be compared with the latest
experimental data.
In Sec.~\ref{SecLC}, we study the effects of the Littlest Higgs
model on the process $e^+ e^- \to \mu^+ \mu^-$, of which the
dominant contribution is from the heavy photon. We summarize our
results in Sec.~\ref{SecConclusion}.

\section{Littlest Higgs model}
\label{SecLH}

At TeV scale, the Littlest Higgs model is embedded into a
non-linear $\sigma$-model with the coset space of $SU(5)/SO(5)$.
The leading two-derivative term for the sigma field $\Sigma$ is
\beq \label{2dim} \mathcal{L}  _{\Sigma} = \frac{1}{2}
\frac{f^2}{4}
    {\rm Tr} | \mathcal{D}_{\mu} \Sigma |^2 .
\eeq The local gauge symmetries $[SU(2) \otimes U(1)]^2$ is also
assumed, which is clear from the following covariant derivative of
the sigma field: \beq \label{covariantD} \mathcal{D}_\mu \Sigma=
\partial_\mu\Sigma - i \sum_{j=1}^2\left( g_jW_j^a( Q_j^a \Sigma +
\Sigma Q_J^{aT}) + g'_j B_j(Y_j\Sigma + \Sigma Y_j^T) \right).
\eeq The generators of two $SU(2)$'s are
\begin{equation}
Q_1^a = \left( \begin{array}{ccc}
\frac{\sigma^a}{2} & & \\
 & & \\
 & & {\mathbf{0}}_{3\times 3}
\end{array}\right),\qquad
Q_2^a = \left( \begin{array}{ccc}
{\mathbf{0}}_{3\times 3} & & \\
 & & \\
 & & -\frac{\sigma^{a*}}{2}
\end{array}\right),
\end{equation}
and two $U(1)$ generators are \beq Y_1 =
\mathrm{diag}(-3,-3,2,2,2)/10, \quad Y_2 =
\mathrm{diag}(-2,-2,-2,3,3)/10 \,. \eeq

At the scale $\Lambda_S \sim 4 \pi  f$, a symmetric tensor of the
$SU(5)$ global symmetry develops an order $f$ vacuum expectation
value (VEV) of which the direction is into the $\Sigma_0$ given by
\begin{equation}
\label{Sigma0} \Sigma_0 = \left( \begin{array}{ccc}
 & & {\mathbf{1}}_{2 \times 2} \\
 &1 & \\
{\mathbf{1}}_{2 \times 2} & &
\end{array}\right).
\end{equation}
Now the following two symmetry breakings occur:
\begin{itemize}
    \item The global $SU(5)$ symmetry is broken into $SO(5)$,
    which leaves 14 massless Goldstone
bosons: They transform under the electroweak gauge group as a real
singlet ${\mathbf{1}}_0$, a real triplet ${\mathbf{3}}_0$, a
complex doublet ${\mathbf{2}}_{\pm \frac{1}{2}}$, and a complex
triplet ${\mathbf{3}}_{\pm 1}$.
    \item The assumed gauge symmetry
$[SU(2)\otimes U(1)]^2$ is also broken into its diagonal subgroup
$SU(2)_L \otimes U(1)_Y$, identified as the SM gauge group. The
gauge fields $\vec{W}^{' \mu }$ and $B^{'\mu }$ associated with
the broken gauge symmetries become massive by eating the Goldstone
bosons of ${\mathbf{1}}_0$ and ${\mathbf{3}}_0$.
\end{itemize}

The non-linear sigma fields are then parameterized by the
Goldstone fluctuations:
\begin{equation}
\Sigma = \Sigma_0 + \frac{2 i}{f} \left( \begin{array}{ccccc}
\phi^{\dagger} & \frac{h^{\dagger}}{\sqrt{2}} &
{\mathbf{0}}_{2\times
2} \\
\frac{h^{*}}{\sqrt{2}} & 0 & \frac{h}{\sqrt{2}} \\
{\mathbf{0}}_{2\times 2} & \frac{h^{T}}{\sqrt{2}} & \phi
\end{array} \right) + {\cal O}(\frac{1}{f^2}),
\end{equation}
where $h$ is a doublet and $\phi$ is a triplet under the unbroken
$SU(2)$. A brief comment is that this Higgs triplet, developing a
non-zero VEV, may explain neutrino mass terms through its Yukawa
coupling with leptons in a SM gauge invariant
way\,\cite{neutrino}. Note that the lepton Yukawa coupling has
some freedom since it is insensitive to the quadratic divergence
of the Higgs if the cutoff scale is around 10 TeV.

The gauge fields $\vec{W}'$ and $B'$ associated with the broken
gauge symmetries are related with the SM gauge fields by
\begin{eqnarray}
    W = s W_1 + c W_2, &\qquad&
    W^{\prime} = -c W_1 + s W_2, \nonumber \\
    B = s^{\prime} B_1 + c^{\prime} B_2, &\qquad&
    B^{\prime} = -c^{\prime} B_1 + s^{\prime} B_2,
\end{eqnarray}
with the mixing angles of \beq
    c = \frac{g_1}{\sqrt{g_1^2+g_2^2}}, \qquad
    c^{\prime} = \frac{g_1^{\prime}}{\sqrt{g_1^{\prime 2}+g_2^{\prime 2}}}.
\eeq The SM gauge couplings are then $g = g_1 s=g_2 c$ and
$g^{\prime} = g_1^{\prime} s^{\prime}=g_2^{\prime} c^{\prime}$. At
the scale $f$, the SM gauge fields remain massless, and the heavy
gauge bosons are massive: \beq \label{MassH} m_{W^{\prime}}
    = \frac{g}{2sc} f , \qquad
    m_{B^{\prime}}
    = \frac{g^{\prime}}{2\sqrt{5}s^{\prime}c^{\prime}}f.
\eeq As shall be discussed in detail, the presence of $\sqrt{5}$
in the denominator of $m_{B^{\prime}}$ implies relatively light
new neutral gauge boson. It is to be compared with the
$SU(6)/Sp(6)$ case of $m_{B^{\prime}}
    = {g^{\prime}f}/({2\sqrt{2}s^{\prime}c^{\prime}})$.

Even though the Higgs boson at tree level remains massless as a
Goldstone boson, its mass is radiatively generated because any
non-linearly realized symmetry is broken by the gauge, Yukawa, and
self interactions of the Higgs field. Early attempts in
constructing a pseudo Goldstone Higgs boson suffered from the same
quadratic divergence as in the SM. Recent little Higgs models
introduce a collective symmetry breaking: Only when multiple gauge
symmetries are broken, the Higgs mass is radiatively generated;
naturally the Higgs mass loop correction occurs at least at two
loop level. In the phenomenological point of view, the
cancellation of the SM contributions at one-loop level occurs as
in the supersymmetry model. For example, the quadratic divergence
due to the SM gauge boson $B^\mu$ is cancelled by that of the
$B^{\prime \mu}$ field, as can be seen
from  
\beq
    \mathcal{L}_{\Sigma}(B\cdot B) \supset
    g^{\prime 2} B_{\mu} B^{\mu}
    {\rm Tr} \left[ \frac{1}{4} h^{\dagger} h
        \right]  -
     g^{\prime 2}
    B_{\mu}^{\prime} B^{\prime \mu}
    {\rm Tr} \left[ \frac{1}{4} h^{\dagger} h \right]
    .
\label{bbhh} \eeq It is clear that the cancellations in little
Higgs models are due to the exactly opposite coupling strength,
which is provided by a larger symmetry structure. It is to be
compared with supersymmetry models where the cancellation occurs
due to opposite spin-statistics between the SM particle and its
super-partner.

Since more severe quadratic divergence of the quantum correction
to the Higgs mass comes from the top quark loop, another
top-quark-like fermion is also required. In addition, this new
fermion is naturally expected to be heavy with mass of order $f$.
We introduce a vector-like fermion pair $\tilde{t}$ and
$\tilde{t}^{\prime c}$ with the SM quantum numbers
$({\mathbf{3,1}})_{Y_i}$ and $({\mathbf{\bar{3},1}})_{-Y_i}$. With
$\chi_i=(b_3, t_3, \tilde{t})$ and antisymmetric tensors of
$\epsilon_{ijk}$ and $\epsilon_{xy}$, the following Yukawa
interaction is chosen in the Littlest Higgs model:
\begin{eqnarray}
{\mathcal{L}}_Y &=& {1\over 2}\lambda_1 f \sum_{i,j,\,k=1}^3\;
\sum_{x,y=4}^5 \epsilon_{ijk} \epsilon_{xy} \chi_i \Sigma_{jx}
\Sigma_{ky} u^{\prime c}_3 \nonumber \\
&+ &\lambda_2 f \tilde{t} \tilde{t}^{\prime c} + {\rm h.c.}
\label{yuk} \nonumber \\
&\supset& -i \lambda_1 ( \sqrt{2} h^0 t_3 + i f \tilde t
        - \frac{i}{f}  h^0h^{0*} \tilde t ) u_3^{\prime c}
            + {\rm h.c.}
            \,.
\label{yuk-expand}
\end{eqnarray}
As Eq.~(\ref{yuk-expand}) shows, the quadratic divergence due to the
 heavy top quark
cancels that due to the SM top quark. And this cancellation is
stable from radiative corrections.

Electroweak symmetry breaking is induced by the remaining
Goldstone bosons $h$ and $\phi$. Through radiative corrections,
the gauge, Yukawa, and self-interaction of the Higgs field
generate a Higgs potential\,\cite{cw}. As discussed before, the
one-loop quadratic divergence of the Higgs mass coefficient
$-\mu^2$ vanishes due to the cancellations between the new gauge
boson (top quark) contributions and the SM contributions. Since
the $\mu^2$ has the log-divergent one-loop and quadratically
divergent two-loop contributions suppressed by a loop factor
$1/16\pi^2$, it is to be treated as a free parameter of order
$100$ GeV. For positive $\mu^2$, the $h$ and $\phi$ fields can
develop VEVs of $\langle h^0 \rangle = v/\sqrt{2}$ and $\langle
\phi^0  \rangle = v'$, which trigger the electroweak symmetry
breaking. Now the SM $W$ and $Z$ bosons acquire masses of order
$v$, and small (of order $v^2/f^2$) mixing between $W$ and $W'$
($Z$ and $Z'$) occurs. In the following, we denote the mass
eigenstates of the SM gauge fields by $W_L$ and $Z_L$.

Some discussions on phenomenological points are in order here.
First, the requirement of positive mass squared of the Higgs
triplet constrains $v'$ to be $v'/v < v/(4f)$. Second, since the
final $U(1)_{\mathrm{QED}}$ symmetry remains intact, the mass and
couplings of photon are the same as in the SM. For the Yukawa
interaction of the other light SM fermions, we assume that
Eq.~(\ref{yuk}) is valid for all SM fermions including leptons,
except that their corresponding extra vector-like fermions are
absent. Then the gauge invariance of Eq.~(\ref{yuk}) with the
anomaly free condition fixes the $U(1)_{1,2}$ charges of the SM
fermions. For example, the lepton doublet and singlet have the
following $U(1)_{1,2}$ charges: \beq \label{U1charge} L  :  Y_1
=-\frac{3}{10}, \qquad Y_2 =-\frac{1}{5}, \qquad e^c : Y_1
=\frac{3}{5}, \qquad Y_2 =\frac{2}{5}. \eeq

The question of the corrections to the electroweak precision data
merits some discussions. The absence of custodial $SU(2)$ global
symmetry in this model yields weak isospin violating contributions
to the electroweak precision observables. In the early study,
global fits to the experimental data put rather severe constraint
on the $f>4$ TeV at 95\% C.L.\cite{p1,p2}. However, their analyses
are based on a simple assumption that the SM fermions are charged
only under one $U(1)$. If all the SM fermions have common Yukawa
couplings with anomaly-free condition as in Eqs.~(\ref{yuk}) and
(\ref{U1charge}), the bounds become relaxed: Substantial parameter
space allows $f\simeq 1-2$ TeV\,\cite{p3,p4}. The experimental
constraints can be more loosed, i.e., by gauging only $U(1)_Y$
from the beginning\,\cite{nelson,p3}. Even though the abandonment
of the one-loop quadratic divergences to the Higgs mass from
$U(1)_Y$ is obviously a theoretical drawback, the resulting
fine-tuning, about 50\% for $\Lambda \sim 10$ TeV, is tolerable.

To illustrate the preferred parameter space consistent with the
low energy data, we present the SM $Z$ boson mass: \beq \label{MZ}
     M_{Z_L}^2 = m_z^2
    \left[ 1 + \Delta \left( \frac{1}{4}
    + c^2 (1-c^2)
    - \frac{5}{4} (c^{\prime 2}-s^{\prime 2})^2 \right)
    + 8 \Delta' \right]
     ,
\eeq where $m_z=gv/(2 c_W)$, $\Delta = v^2/f^2 \ll 1 $, $\Delta' =
v'^2/v^2 \ll 1$, $t_W =s_W/c_W$ and \beq \nonumber x_H =g
g^{\prime}
    \frac{scs^{\prime}c^{\prime} (c^2s^{\prime 2} + s^2c^{\prime 2})}
    {2g^2 s^{\prime 2} c^{\prime 2} - 2g^{\prime 2} s^2 c^2/5}
    .
\eeq And the gauge couplings of the $Z_L$ with the charged
leptons, in the form of $\, \gamma^\mu (g_L^Z P_L + g_R^Z P_R)$
with $P_{R,L}=(1 \pm \gamma^5)/2$, are \bea \label{gZ} g_R^{Z} &=&
\frac{e}{s_W c_W} \left[ -\frac{1}{2}+s_W^2 +\Delta \left\{
\frac{c^2}{2}(c^2-\frac{1}{2})
-\frac{5}{4}(c'^2-s'^2)(c'^2-\frac{2}{5}) \right\} \right] ,
\\
\nonumber g_L^Z &=& \frac{e}{s_W c_W} \left[ s_W^2
+\frac{5}{2}\Delta (c'^2-s'^2)(c'^2-\frac{2}{5}) \right] \,. \eea
Here the QED bare coupling $e^2$ is the running coupling at the
$Z$-pole, and the bare value of $s_W^2$ is related with the
measure value of $s_0^2$ by \beq \label{invswcw} \frac{1}{s_W c_W}
=\frac{1}{s_0 c_0} \left[ 1-\frac{\Delta}{2} \left\{ c^2 s^2
-\frac{5}{4}(c'^2-s'^2)^2 \right\} -2 \Delta' \right] \,. \eeq
Since the main corrections to low energy observables in
Eqs.~(\ref{MZ})-(\ref{invswcw}) are proportional to $c^2$ or
$(c'^2-s'^2)$, the parameter space around $c \ll 1$ and
$c'=1/\sqrt{2}$ suppresses new contributions: The $f$ bound about
2 TeV is allowed in the region around $c \in [0, 0.5 ]$ and $c'
\in [0.62, 0.73]$\,\cite{p3}.

\section{Properties of the heavy photon}
\label{SecAH}

Among various little Higgs models such as the $SU(6)/Sp(6)$ model,
the $SU(4)^4/SU(3)^3$ model, the $SO(5)^8/SO(5)^4$ model and so
on, the Littlest Higgs model can be distinguished by the presence
of a relatively light $A_H$. From Eq.~(\ref{MassH}), the mass
ratio of the heavy photon $A_H$ to the $Z_H$ is \beq
\frac{M_{A_H}^2}{M_{Z_H}^2} = \frac{s_W^2}{5\,c_W^2}\frac{s^2
c^2}{s'^2 c'^2} + \mathcal{O}\left( \frac{v^2}{f^2} \right) \sim
0.06 \left( \frac{s^2 c^2}{s'^2 c'^2} \right) \,. \eeq Even for
the case of $c\simeq c'$, the $A_H$ is substantially lighter than
the $Z_H (W_H)$. In addition, the electroweak precision data
prefer the parameter space of $c \ll 1$ and $c' \sim 1/\sqrt{2}$,
which much more suppresses the ratio ${M_{A_H}^2}/{M_{Z_H}^2}$.

\begin{figure}[thb]
\begin{center}
\begin{center}
    \includegraphics[scale=1.0]{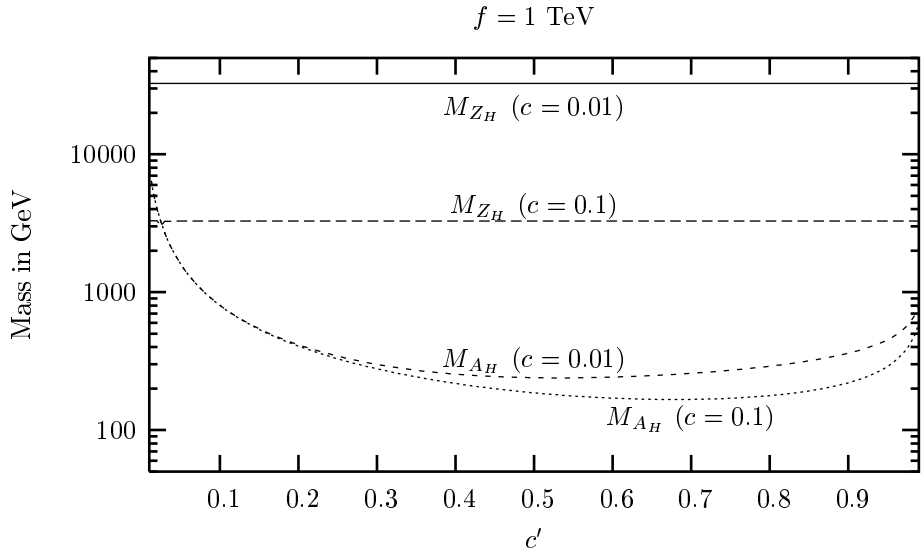}
    \end{center}
    \end{center}
    \caption {The masses of $A_H$ and $Z_H$ in GeV as a function of
    $c'$ for $c=0.01,~0.1$.
    The $f$ is set to be 1 TeV.
    }
    \label{figMAH-MZH}
\end{figure}
In Fig.~\ref{figMAH-MZH}, we present the $M_{A_H}$ and $M_{Z_H}$
as a function of $c'$ with the fixed $f=1$ TeV (the $M_{A_H}$ and
$M_{Z_H}$ increase linearly with $f$). In most of the parameter
space, the $A_H$ is much lighter than the $Z_H$: Around
$c'=1/\sqrt{2}$ and $c=0$, where the corrections to the
electroweak precision data are minimized, the mass difference is
maximized.
Since the heavy photon is mainly the $B'$, 
its mass depends weakly on the value of $c$, the mixing parameter
between two $SU(2)$ gauge bosons. The $M_{A_H}$ for $c=0.3$ is
practically identical with that for $c=0.1$. On the contrary, the
$M_{Z_H}$ is sensitive to $c$, while almost insensitive to $c'$.
In particular, small value of $c$ enhances the mass of $Z_H$: For
$c<0.1$, the $Z_H$ becomes too heavy to be sufficiently produced
at LHC.

Note that in the parameter space consistent with the low energy
observables, the heavy photon becomes light enough to be produced
at the future linear collider.
\begin{figure}[thb]
\begin{center}
    \includegraphics[scale=1.0]{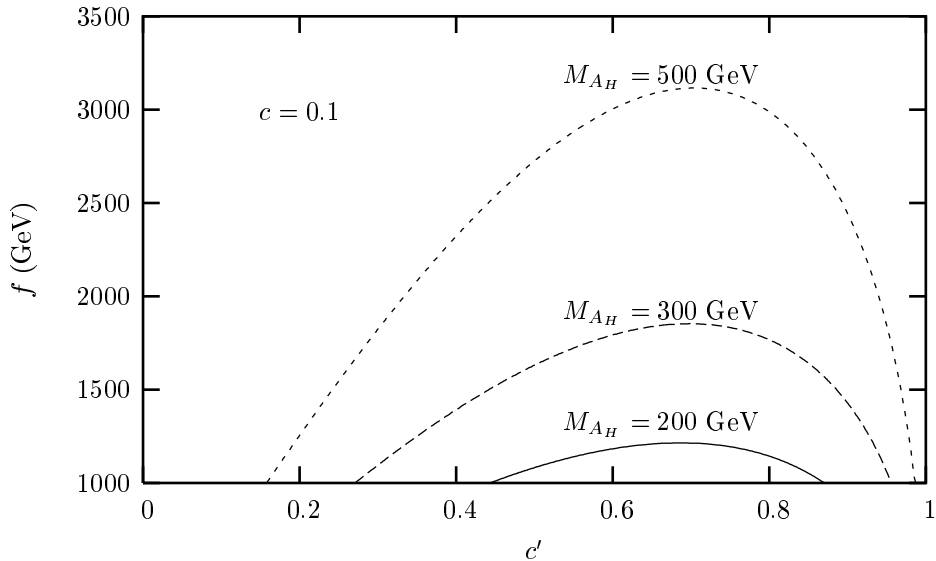}
    \end{center}
    \caption {In the parameter space of $(c',f)$,
    contours for $M_{A_H}=200,~300,~500$ GeV.
    The value of $c$ is fixed to be 0.1.
    }
    \label{figcp-f}
\end{figure}
Figure \ref{figcp-f} illustrates contours for
$M_{A_H}=200,~300,~500$ GeV in the parameter space of $(c',f)$.
The value of $c$, which little affects $M_{A_H}$, is set to be
$0.1$. In particular, the region around $c'=1/\sqrt{2}$ allows the
on-shell production of the heavy photon at 500 GeV linear
colliders for $f \lsim 3$ TeV.

Next the gauge couplings of heavy neutral gauge bosons are \beq
\mathcal{L} = -(c'^2 Y_1 -s'^2 Y_2) \bar{f}\gamma_\mu f A_H^\mu +
\frac{g c}{s}\overline{Q}_L \gamma_\mu T^3 Q_L Z_H^\mu \,. \eeq In
principle, if $s'^2/c'^2=Y_1/Y_2$, which is allowed only when the
anomaly-free condition is violated, the vertex of $f-\bar{f}-A_H$
vanishes. Another crucial point is that the right-handed top quark
coupling with the $A_H$ has the additional term: \beq
\label{top-coupling} g_R^{A_H-t-\bar{t}} = \frac{g'}{ s'c'}
\left(\frac{4}{3} - \frac{5}{6}c'^2 -
\frac{1}{5}\frac{\lambda_1^2}{\lambda_1^2+\lambda_2^2} \right),
\eeq while the left-handed top quark does not have, as shown by
\beq g_L^{A_H-t-\bar{t}} = \frac{g'}{ s'c'} \left(\frac{1}{15} -
\frac{1}{6}c'^2 \right) . \eeq This is attributed to the proposed
top Yukawa coupling in Eq.~(\ref{yuk}). The physical mass
eigenstates of $SU(2)$-singlet top quark $t_R^c$ and heavy top
quark $T_R$ are the mixtures of weak eigenstates, $u^{\prime c}_3$
and $\tilde{t}^{\prime c}$: \beq t_R^c =
\frac{1}{\sqrt{\lambda_1^2+\lambda_2^2}} \left( -\lambda_1
\tilde{t}^{\prime c} +\lambda_2 u^{\prime c}_3 \right) \,,\quad
T_R = \frac{1}{\sqrt{\lambda_1^2+\lambda_2^2}} \left( -\lambda_1
\tilde{t}^{\prime c} +\lambda_2 u^{\prime c}_3 \right) \,, \eeq
which explain the second terms of the $g_{R}^{A_H-t-\bar{t}}$ in
Eq.~(\ref{top-coupling}). Even in the special cases of the
suppressed $A_H$ couplings, the top quark can substantially
interact with the $A_H$.

\begin{figure}[thb]
\begin{center}
    \includegraphics[scale=0.9]{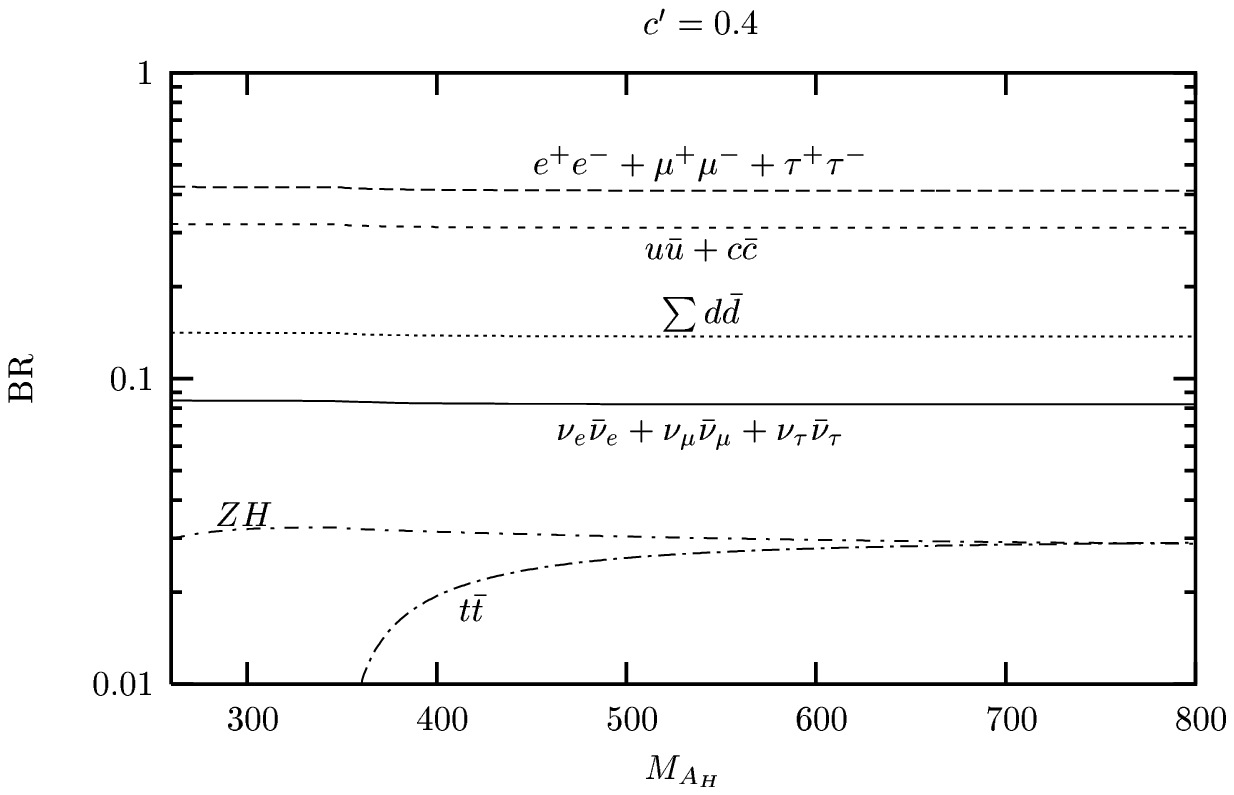}
    \includegraphics[scale=0.9]{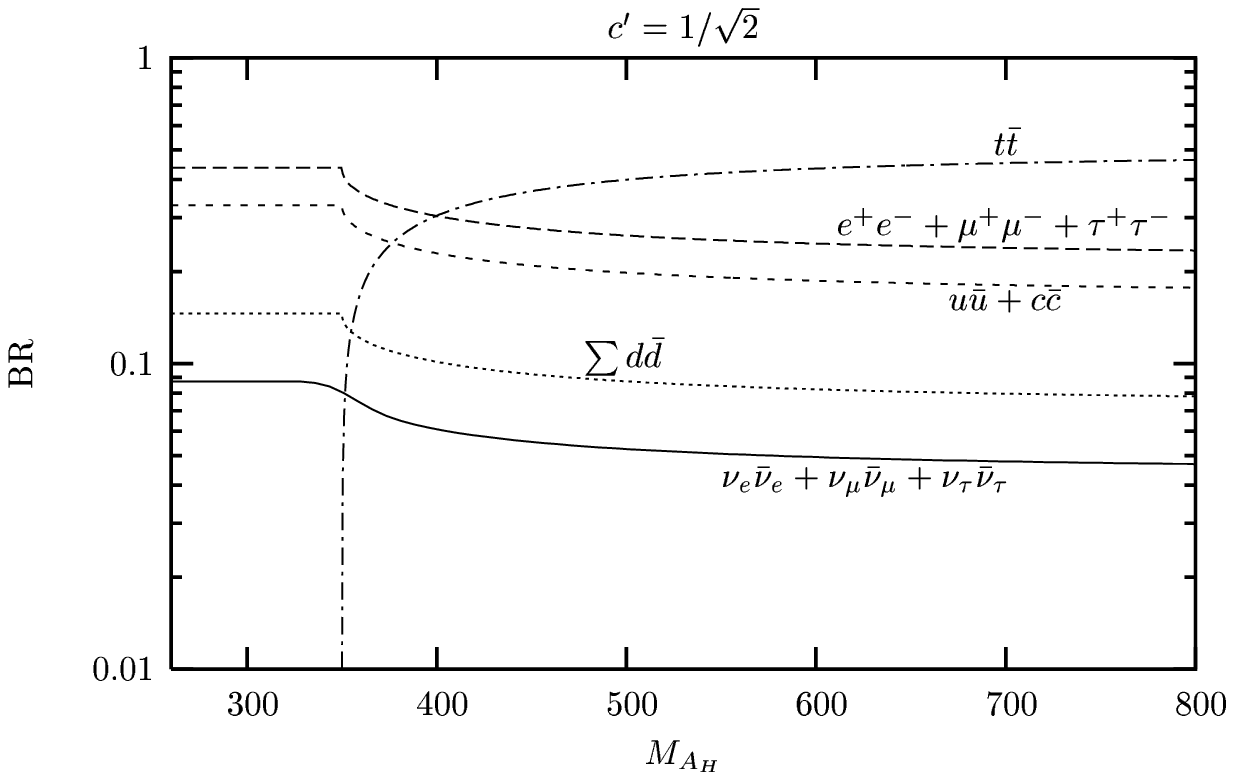}
    \end{center}
    \caption {The branch ratios of the $A_H$
    as a function of $M_{A_H}$
    for $c'=0.4$ and $c'=1/\sqrt{2}$.
    The Higgs mass is 120 GeV.
    }
    \label{figBR}
\end{figure}
Now let us discuss about the decay of the $A_H$ into a fermion
pair and $Z-h$. Decay into a SM $W_L$ pair is suppressed by a
factor of $(v/f)^4$. Partial decay rates are \bea \Gamma(A_H \to f
\bar{f}) &=& \frac{N_c}{12 \pi } \left[ (g_v^{A_H})^2 (1+2 r_f) +
(g_a^{A_H})^2 (1-4 r_f)\right] \sqrt{1-4 r_f} M_{A_H}, \\
\nonumber \Gamma(A_H \to Z h) &=& \frac{g'^2(c'^2-s'^2)}{384 \pi
c's'} \lambda^{1/2} [
 (1+r_Z-r_h)^2+8 r_Z] M_{A_H},
\eea where $N_c$ is the color factor, $r_i =m_i^2/M_{A_H}^2$, and
$\lambda=1+r_Z^2+r_h^2+2r_Z+2 r_h+2 r_Z r_h$. Note that if
$c'=1/\sqrt{2}$, the $A_H$ decay into $Z \,h $ is prohibited. We
refer the reader for the full expressions of the $g_{v,a}^{A_H}$
to Ref.~\cite{Han}. In Fig.~\ref{figBR}, we show the branching
ratios of the $A_H$ as a function of $M_{A_H}$ for $c'=0.4$ and
$c'=1/\sqrt{2}$. Two top quark Yukawa couplings $\lambda_1$ and
$\lambda_2$ in Eq.~(\ref{yuk}) are assumed to be equal to each
other. Except for the narrow region around $c'=1/\sqrt{2}$, the BR
patterns are almost the same: The decay into a charged lepton pair
is dominant. If $c'=1/\sqrt{2}$, the $A_H-Z-h$ coupling vanishes
and the $A_H$ gauge couplings with a lepton pair is suppressed
since then $s'^2/c'^2=1$ is close to $Y_1/Y_2 =3/2$. In this case,
the decay into top quark become dominant if kinematically allowed.

\begin{figure}[thb]
\begin{center}
    \includegraphics[scale=1.0]{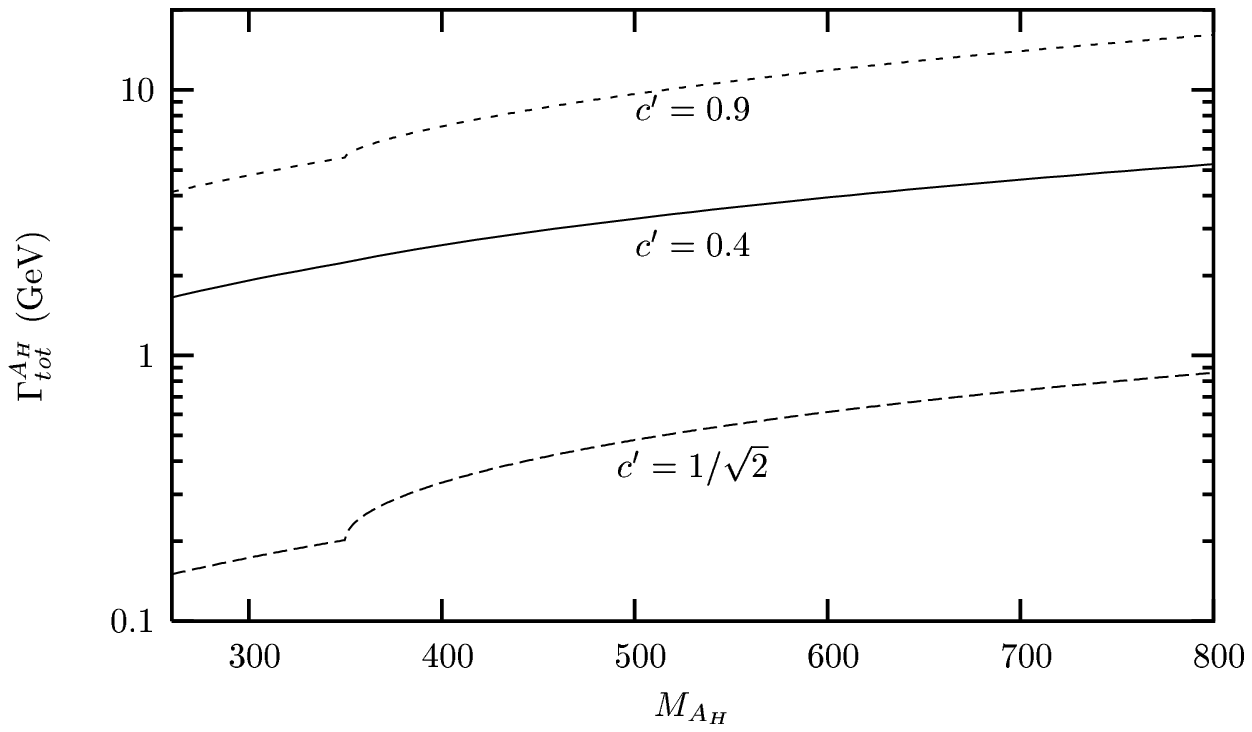}
\end{center}
    \caption {The total decay rate of the $A_H$
    as a function of $M_{A_H}$
    for $c'=0.4$ and $c'=1/\sqrt{2}$.
    The Higgs mass is 120 GeV.
    }
    \label{figGam}
\end{figure}
In Fig.~\ref{figGam}, we show the total decay rate of the $A_H$ as
a function of $M_{A_H}$ for $c'=0.4,~1/\sqrt{2},~0.9$. In
particular, the $c'=1/\sqrt{2}$ case yields a very narrow
resonance peak, raising a possibility that the resonance peak
might be missed.

The next question is whether this light $A_H$ is consistent with
the recent measurement of muon anomalous magnetic
moment\,\cite{Bennett:2002jb}(see also recent review in
Ref.~\cite{Gray:2003fc}). This issue shall be answered in the
following section.

\section{Anomalous magnetic moment of muon and little Higgs }
\label{SecMuon}
%

 In this
section, we study the one-loop level contribution of the Littlest
Higgs model by calculating its effects on the anomalous magnetic
moment of muon. The present status of the theoretical evaluation
of the $(g-2)_\mu$ in the SM is not conclusive because of the
inconsistent values between the hadronic vacuum polarizations
based on $e^+ e^-$ and $\tau$ data. Comparison with the
experimental value implies \bea \label{expg2} a_\mu^{\mathrm{exp}}
- a_\mu^{\mathrm{SM}}(e^+ e^-) &=& (35.5 \pm 11.7) \times 10^{-10}
\qquad [\,3\,\sigma],
 \\ \nonumber
a_\mu^{\mathrm{exp}} - a_\mu^{\mathrm{SM}}(\tau) &=& (10.3 \pm
10.7) \times 10^{-10}\qquad
 [\,1\,\sigma]
\,. \eea

\begin{figure}[thb]
    \begin{center}
    \includegraphics[scale=1.0]{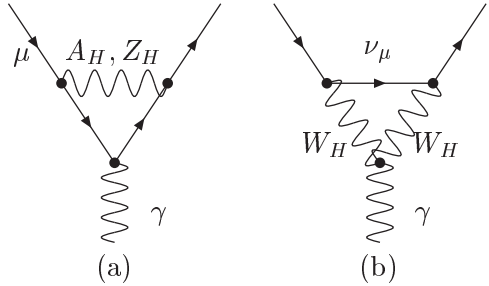}
    \end{center}
    \caption {Feynman diagrams for one-loop contribution of
        heavy gauge bosons.
(a) shows the contributions from $A_H, Z_H$
        and (b) shows the contribution from $W_H$.
    }
    \label{Feynman diagram}
\end{figure}
In the Littlest Higgs model, one-loop corrections to the muon
anomalous magnetic moment come from the Feynman diagrams mediated
by the heavy photon $A_H$, the new heavy neutral weak boson $Z_H$
and the new heavy charged weak boson $W_H$ as depicted in
Fig.\ref{Feynman diagram}. Since each contribution to $\Delta
a_\mu$ is inversely proportional to the gauge boson mass squared
and the $M_{A_H}^2$ is smaller than the $M^2_{Z_H,W_H}$ at least
by an order of magnitude, we consider only the $A_H$ contribution
of\,\cite{moore} \beq \label{g-2} \Delta a_{A_H}
=\frac{1}{12\pi^2} \left(\frac{m_\mu}{M_{A_H}^2}\right)^2
\left[(g_v^{A_H})^2 - 5 (g_a^{A_H})^2\right] \,. \eeq

\begin{figure}[h]
\begin{center}
    \includegraphics[scale=1.0]{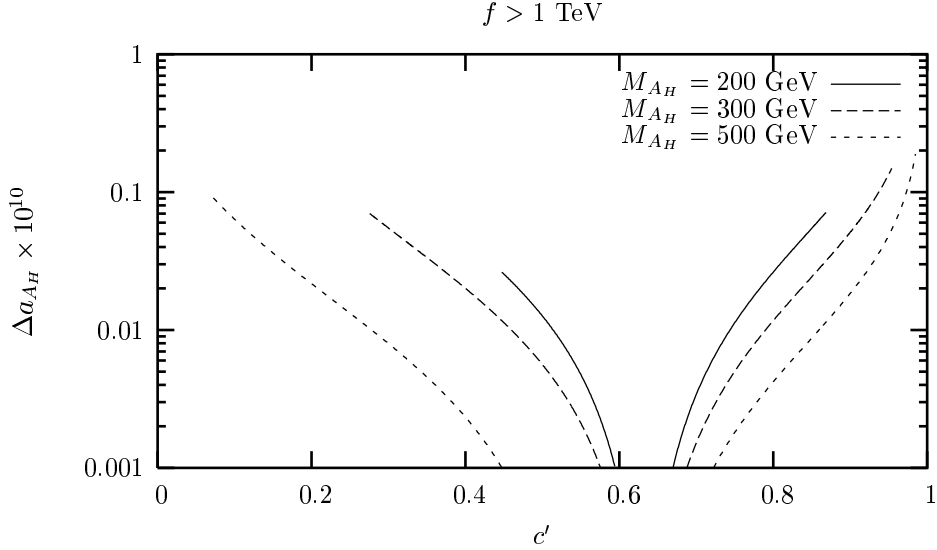}
\end{center}
    \caption {The contribution to the muon anomalous
    magnetic moments due to the heavy photon
    as a function of $c'$
    for $M_{A_H}=200,~300,~500$ GeV.
    }
    \label{fig-g2}
\end{figure}
Figure \ref{fig-g2} shows the $\Delta a_{A_H}$ as a function of
$c'$ for fixed $M_{A_H}=200,~300,~500$ GeV. Since we require $f>1$
TeV, limited space of $c'$ according to the $M_{A_H}$ is
presented. The contribution to the $\Delta a_\mu$ increases as the
$M_{A_H}$ decreases and the $c'$ deviates from the value of
$s'^2/c'^2 =Y_1/Y_2$. In the whole parameter space, the $\Delta
a_{A_H}$ is quite safe from the recent experimental data in
Eq.~(\ref{expg2}). Therefore, it is concluded that the light $A_H$
is not inconsistent with the current status of the theoretical and
experimental value of the muon anomalous magnetic moment.

\section{$e^+ e^- \to \mu^+ \mu^-$}
\label{SecLC} In the Littlest Higgs model the heavy photon mass
$M_{A_H}$ depends on the global symmetry breaking scale $f$, two
mixing angles of $c$ and $c'$ between light and heavy gauge
bosons. In Sec.~\ref{SecAH}, we have shown that as the model
parameters are arranged to suppress the contributions to the
electroweak precision data, this $M_{A_H}$ reduces, which is
contrary to the usual case where the heavier new particle
suppresses the corrections to the low energy observables. And in a
major portion of parameter space, the dominant decay mode is shown
to be into a charged lepton pair. Therefore, one of the most
effective signals to probe the model can come from the process
$e^+ e^- \to \mu^+ \mu^-$.

The process $e^+ e^- \to \mu^+ \mu^-$ has two SM $s$-channel
Feynman diagrams mediated by the photon and $Z$ boson. In the
Littlest Higgs model, it has two additional $s$-channel diagrams
mediated by the $A_H$ and $Z_H$. The corresponding helicity
amplitude $\M _{\lm_e \bar{\lm}_e \lm_\mu \bar{\lm}_\mu }$, where
the $\lm_l$ and $\bar{\lm}_l$ are respectively the polarization of
$l^-$ and $l^+$, can be simplified by $\M _{\lm_e\lm_\mu}$ since
$\lm_l = - \bar{\lm}_l$ with the lepton mass neglected. We have
\beq \M _{\lm_e\lm_\mu} = -(1+\lm_e \lm_\mu \cos\theta) \sum_{V_j}
g_{\lm_e}^{V_j} g_{\lm_e}^{V_j} \mathcal{D}_{V_j} \,, \eeq where
$\theta$ is the scattering angle of the muon with respect to the
electron beam, $V_j=A,Z,A_H,Z_H$, and the $\mathcal{D}_{V_j}$ is
the propagation factor of \beq \mathcal{D}_{V_j} =
\frac{s}{s-M_{V_j}^2+i M_{V_j} \Gamma_{V_j}} \,. \eeq And the
$g_{\lm_e}^{V_j} \left(= g_{\lm_e}^{V_j-l^+ - l^-}\right)$'s are
\bea \label{gZH} g_R^{Z_H} &=& -\frac{g c }{2 s} ,
\qquad\qquad\qquad\qquad
g_L^{Z_H} = 0 \\
\label{gAH} g_R^{A_H} &=& \frac{g'}{s'c'} \left( -\frac{1}{5}
+\frac{c'^2}{2} \right), \qquad g_L^{A_H} = \frac{g'}{s'c'} \left(
-\frac{2}{5} +{c'^2} \right) \,. \eea

\begin{figure}[thb]
    \includegraphics[scale=1.0]{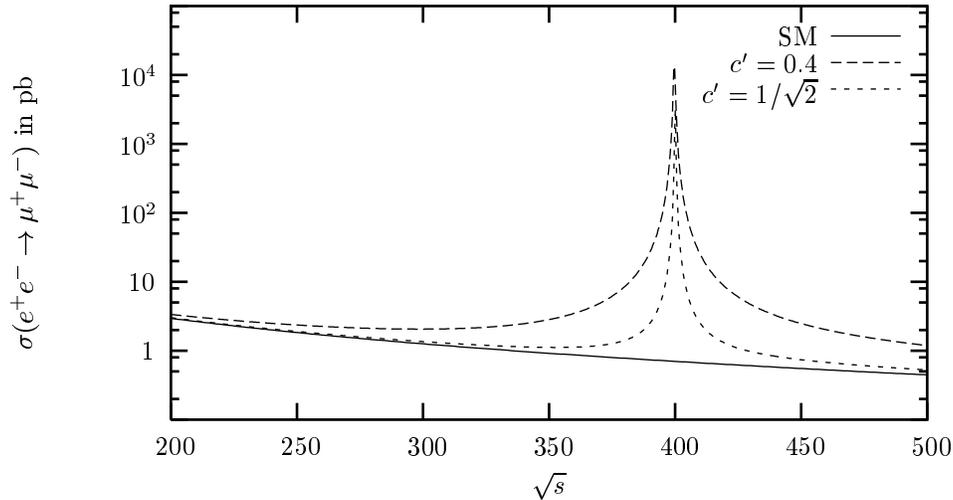}
    \caption {The total cross section of the process
as a function of $\sqrt{s}$ with $M_{A_H}=400$ GeV and $c=0.1$. We
consider two cases of $c'=0.4$ and $c'=1/\sqrt{2}$.
    }
\label{figtotsg}
\end{figure}
In Fig.~\ref{figtotsg}, we present the total cross section as a
function of $\sqrt{s}$ for $c'=0.4$ and $c'=1/\sqrt{2}$. We set
$M_{A_H}=400$ GeV and $c=0.1$. If the $c'$ sizably deviates from
the critical point of $s'^2/c'^2=Y_1/Y_2$ (see $c'=0.4$ case), the
coupling $A_H - l^+ - l^-$ is large enough to yield substantial
deviation from the SM results even outside the resonance peak. In
the parameter region of the suppressed $A_H - l^+ - l^-$ coupling
(see $c'=1/\sqrt{2}$ case), only around the resonance peak can
produce significant new signal.

\section{Summary and Conclusion}
\label{SecConclusion}

The Littlest Higgs model could be an alternative model for new
physics beyond the Standard Model which solves the little
hierarchy problem. From the extension in the gauge sector, we
expect a new set of gauge bosons. The heavy photon is shown to be
lightest of all and, moreover, lighter in the preferred parameter
space by electroweak precision measurements. We checked the
consistency of the parameter space by calculating one-loop induced
anomalous magnetic moment of muon. Its numerical value is $\Delta
a_\mu \leq 0.1 \times 10^{-10}$ in the whole parameter region.
Then we study the on-shell production and decay of the heavy
photon in the future linear collider. The heavy photon mainly
decays to lepton pairs but for $c'=1/\sqrt{2}$ it could mainly
decay to a top-quark pair if the kinematics allows. The high
energy process of $e^+ e^- \rightarrow \mu^+ \mu^-$ is explicitly
considered and the resonance structure of the heavy photon
production is shown for various parameters of the model. In
conclusion, the heavy photon of the Littlest Higgs induces
negligible corrections to the anomalous magnetic moment of muon
but still can be produced in the future linear collider.

\acknowledgments
Authors thank to Prof. J.~E.~Kim, K.~Choi and
E.~J.~Chun for valuable comments on our work. SC also thanks to
Prof. T.~Moroi and M.~Yamaguchi for valuable discussions during
the KIAS-KAIST joint workshop.
The work of JS is supported by the faculty research fund of Konkuk
University in 2003, and partly by
Grant No. R02-2003-000-10050-0 from BR of the KOSEF.

\end{document}